\documentclass[prd,aps,floats,nofootinbib,showpacs,preprintnumbers]{revtex4}
\usepackage{epsfig}

\newcommand{\be}{\begin{equation}}
\newcommand{\ee}{\end{equation}}
\newcommand{\bea}{\begin{eqnarray}}
\newcommand{\eea}{\end{eqnarray}}
\newcommand{\gag}{g_{a\gamma}}

\begin{document}

\title{Photon-axion conversion in intergalactic magnetic fields
and cosmological consequences\footnote{Contribution  to appear in
  a volume of Lecture Notes in Physics (Springer-Verlag) on Axions.}
}

\author{       Alessandro Mirizzi}
\affiliation{   Dipartimento di Fisica
                and Sezione INFN di Bari\\
                Via Amendola 173, 70126 Bari, Italy\\}

\author{       Georg G.~Raffelt}
\affiliation    {   Max-Planck-Institut f\"{u}r Physik
                (Werner-Heisenberg-Institut)\\
                F\"{o}hringer Ring 6, 80805 M\"{u}nchen, Germany}

\author{       Pasquale D.~Serpico}
\affiliation    {   Max-Planck-Institut f\"{u}r Physik
                (Werner-Heisenberg-Institut)\\
                F\"{o}hringer Ring 6, 80805 M\"{u}nchen, Germany}

%%%%%%%%%%%%%%%%%%%%%%%%%%%%%%%%%%%%%%%%%%%%%%%%%%%%%%%%%%%%%%%%%%%%%%
\begin{abstract}
%%%%%%%%%%%%%%%%%%%%%%%%%%%%%%%%%%%%%%%%%%%%%%%%%%%%%%%%%%%%%%%%%%%%%%
Photon-axion conversion induced by intergalactic magnetic fields
causes an apparent dimming of distant sources, notably of cosmic
standard candles such as supernovae of type Ia (SNe~Ia). We review the
impact of this mechanism on the luminosity-redshift relation of
SNe~Ia, on the dispersion of quasar spectra, and on the spectrum of
the cosmic microwave background.  The original idea of explaining the
apparent dimming of distant SNe~Ia without cosmic acceleration is
strongly constrained by these arguments. However, the cosmic equation
of state extracted from the SN~Ia luminosity-redshift relation remains
sensitive to this mechanism.  For example, it can mimic phantom
energy.

\end{abstract}
%%%%%%%%%%%%%%%%%%%%%%%%%%%%%%%%%%%%%%%%%%%%%%%%%%%%%%%%%%%%%%%%%%%%%%
%%%%%%%%%%%%%%%%%%%%%%%%%%%%%%%%%%%%%%%%%%%%%%%%%%%%%%%%%%%%%%%%%%%%%%
\pacs{98.80.Es, % Observational cosmology
98.80.Cq, %Particle-theory and field-theory models of the early Universe
14.80.Mz. %Axions and other Nambu-Goldstone bosons (Majorons, familons, etc.)
\hfill Preprint MPP-2006-87}

\maketitle

\newpage
%%%%%%%%%%%%%%%%%%%%%%%%%%%%%%%%%%%%%%%%%%%%%%%%%%%%%%%%%%%%%%%%%%%%%%
\section{Introduction}

The two-photon coupling of axions or axion-like particles allows for
transitions between them and photons in the presence of external
electric or magnetic fields as shown in Figure~1~\cite{sikivie,
Raffelt:1987im}. This mechanism serves as the basis for the
experimental searches for galactic dark matter axions~\cite{sikivie,
Bradley:2003kg} and solar axions~\cite{sikivie, vanBibber:1988ge,
Moriyama:1998kd, Inoue:2002qy, Andriamonje:2004hi}. The
astrophysical implications of this mechanism have also been widely
investigated and reviewed~\cite{Raffeltbook, Raffelt:1999tx}. The
phenomenological consequences of an extremely light or massless
axion would be particularly interesting in several astrophysical
circumstances such as polarization of
radio-galaxies~\cite{Harari:1992ea} and
quasars~\cite{Hutsemekers:2005iz}, the diffuse x-ray
background~\cite{Krasnikov:1996bm}, or ultra-high energy cosmic
rays~\cite{Gorbunov:2001gc, Csaki:2003ef}.

One intriguing cosmological consequence of this mechanism is
photon-axion conversion caused by intergalactic magnetic fields,
leading to the dimming of distant light sources, notably of
supernovae (SNe) of type Ia that are used as cosmic standard
candles~\cite{Csaki:2001yk}. Observationally, SNe~Ia at redshifts
$0.3 \lesssim z \lesssim 1.7$ appear fainter than expected from the
luminosity-redshift relation in a decelerating
Universe~\cite{supnovR, supnovP, Riess:2004nr}, a finding usually
interpreted as evidence for acceleration of the cosmic expansion
rate and thus for a cosmic equation of state (EoS) that today is
dominated by a cosmological constant, a slowly evolving scalar
field, or something yet more exotic~\cite{Carroll:2003qq}. The
dimming caused by photon-axion conversion could mimic this behavior
and thus provide an alternative to the interpretation as cosmic
acceleration. Although still requiring some non-standard fluid to
fit the flatness of the universe, this model seemed capable to
explain the SN dimming through a completely different mechanism.

However, if the light from distant SNe~Ia reaches us partially
converted to axion-like particles, the same mechanism would affect
all distant sources of electromagnetic radiation. Therefore, it
appears useful to update the different arguments constraining
photon-axion conversion in intergalactic magnetic fields, in
particular the constraints arising from spectral distortions of the
cosmic microwave background (CMB) and dispersion of quasar (QSO)
spectra.

To this end we begin in Section~2 with a review of the formalism of
photon-axion conversion in magnetic fields. Some technical details are
deferred to Appendix~A. In Section~3 we describe how this mechanism
affects the SN~Ia luminosity-redshift relation and accounts for the
observed dimming. In Section~4 we turn to spectral CMB distortions and
in Section~\ref{QSO} combine these limits with those from the
dispersion of QSO spectra. In Section~6 we describe some additional
limits from a violation of the reciprocity relation between the
luminosity and angular diameter distances. We conclude in
Section~\ref{conclusions} and comment on the viability of the
photon-axion conversion mechanism.

%%%%%%%%%%%%%%%%%%FIG01%%%%%%%%%%%%%%%%%%%%%%%%%%%%%%%%%%%%%%%%%%%%%%%
\begin{figure}[b]
\centering
\includegraphics[height=3cm]{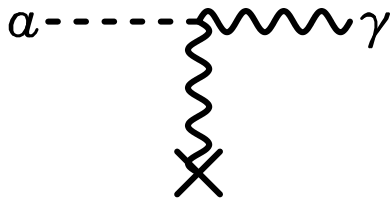}
\caption{ \footnotesize\baselineskip=4mm
Axion-photon transition in an external electric or magnetic
field.}
\label{fig:1}       % Give a unique label
\end{figure}
%%%%%%%%%%%%%%%%%%%%%%%%%%%%%%%%%%%%%%%%%%%%%%%%%%%%%%%%%%%%%%%%%%%%%%

%%%%%%%%%%%%%%%%%%%%%%%%%%%%%%%%%%%%%%%%%%%%%%%%%%%%%%%%%%%%%%%%%%%%%%
\section{Photon-axion conversion}
\label{sec:conversion}

To understand how photon-axion conversion could affect distant sources
we take a closer look at the phenomenon of photon-axion mixing.  The
Lagrangian describing the photon-axion system is~\cite{Raffeltbook}
%%%%%%%%%%%%%%%%%%%%%%%%%%%%%%%%%%%%%%%%%%%%%%%%%%%%%%5
\begin{equation}
{\cal L} = {\cal L}_{\gamma} + {\cal L}_a + {\cal L}_{a \gamma} \,\ .
\end{equation}
%%%%%%%%%%%%%%%%%%%%%%%%%%%%%%%%%%%%%%%%%%%%%%%%%%%%%%
The QED Lagrangian for photons is
%%%%%%%%%%%%%%%%%%%%%%%%%%%%%%%%%%%%%%%%%%%%%%%%%%%%%
\begin{equation}
{\cal L}_{\gamma} = - \frac{1}{4} F_{\mu\nu}{F}^{\mu\nu} +
\frac{\alpha^2}{90 m_e^4}\left[(F_{\mu\nu}{F}^{\mu\nu})^2
+ \frac{7}{4} (F_{\mu\nu}\tilde{F}^{\mu\nu})^2 \right] \,\ ,
\end{equation}
%%%%%%%%%%%%%%%%%%%%%%%%%%%%%%%%%%%%%%%%%%%%%%%%%%%%%
where $F_{\mu\nu}$ is the electromagnetic field-strength tensor,
$\tilde{F}_{\mu\nu}
=\frac{1}{2}\epsilon_{\mu\nu\rho\sigma}F^{\rho\sigma}$ its dual,
$\alpha$ the fine-structure constant, and $m_e$ the electron mass.
We always use natural units with $\hbar=c=k_{\rm B}=1$.  The second
term on the r.h.s. is the Euler-Heisenberg effective Lagrangian,
describing the one-loop corrections to classical electrodynamics for
photon frequencies $\omega\ll m_e$.

The Lagrangian for the non-interacting axion field $a$ is
%%%%%%%%%%%%%%%%%%%%%%%%%%%%%%%%%%%%%%%%%%%%%%%%%%%%%%%%
\begin{equation}
 {\cal L}_{a} = \frac{1}{2} \partial^{\mu} a \partial_{\mu}a
 -\frac{1}{2} m^2 a^2 \,\ .
 \end{equation}
%%%%%%%%%%%%%%%%%%%%%%%%%%%%%%%%%%%%%%%%%%%%%%%%%%%%%%
A generic feature of axion models is the CP-conserving two-photon
coupling, so that the axion-photon interaction is
%%%%%%%%%%%%%%%%%%%%%%%%%%%%%%%%%%%%%%%%%%%%%%%%%%%%%%%%%%%%%%%%%%%%%
\begin{equation}
{\cal L}_{a \gamma}=  -\frac{1}{4}\,\gag
F_{\mu\nu}\tilde{F}^{\mu\nu}a  =  \gag \,\ {\bf E}\cdot{\bf B}\,a ,
\end{equation}
%%%%%%%%%%%%%%%%%%%%%%%%%%%%%%%%%%%%%%%%%%%%%%%%%%%%%%%%%%%%%%%%%%%%%%
where $\gag$ is the axion-photon coupling with dimension of inverse
energy.  A crucial consequence of ${\cal L}$ is that the propagation
eigenstates of the photon-axion system differ from the corresponding
interaction eigenstates. Hence interconversion takes place, much in
the same way as it happens for massive neutrinos of different
flavors. However, since the mixing term
$F_{\mu\nu}\tilde{F}^{\mu\nu}a$ involves two photons, one of them must
correspond to an external field~\cite{sikivie, Raffelt:1987im,
Raffeltbook, Anselm:1987vj}.

Axion-photon oscillations are described by the coupled Klein-Gordon
and Maxwell equations implied by these Lagrangians.  For very
relativistic axions ($m_a \ll \omega$), the short-wavelength
approximation can be applied and the equations of motion reduce to a
first order propagation equation.  More specifically, we
consider a monochromatic light beam travelling along the
$z$-direction in the presence of an arbitrary magnetic field
$\bf{B}$. Accordingly, the propagation equation takes the
form~\cite{Raffelt:1987im}
\begin{equation}\label{linsys1}
\left(\omega-i\partial_z +{\cal M}\right)
\left(
\begin{array}{ccc}
A_x\\
A_y\\
a
\end{array}
\right)=0\,,
\end{equation}
where  $A_x$ and $A_y$ correspond to the two linear photon
polarization states, and $\omega$ is the photon or axion energy. The
mixing matrix is
\begin{equation}  \label{mixmatxy}
{\cal M}=\left(
  \begin{array}{ccc}
    \Delta_{xx}&\Delta_{xy}&\frac{1}{2}\gag B_x\\
    \Delta_{yx}&\Delta_{yy}&\frac{1}{2}\gag B_y\\
    \frac{1}{2}\gag B_x&\frac{1}{2}\gag B_y& \Delta_a
  \end{array}
  \right)\,,
\end{equation}
where $\Delta_a = -m^2_{a}/2\omega$. The component of ${\bf B}$
parallel to the direction of motion does not induce photon-axion
mixing. While the terms appearing in the third row and column of
 ${\cal M}$ have an evident physical meaning, the
 $\Delta_{ij}$-terms ($i,j=x,y$) require some explanations.
  Generally speaking, they are determined
both by the properties of the medium and the QED vacuum polarization
effect.  We ignore the latter, being sub-dominant for the problem at
hand~\cite{Deffayet:2001pc}.

For a homogeneous magnetic field we may choose the $y$-axis along the
projection of $\bf{B}$ perpendicular to the $z$-axis.  Correspondingly
we have $B_x=0$, $B_y = |\bf{B_{\rm T}}| = B \sin \theta$, $A_x =
A_\perp$, $A_y = A_\parallel$. Equation~(\ref{linsys1}) then becomes
\begin{equation}  \label{linsys}
\left(\omega-i\partial_z +{\cal M}\right)
\left(
\begin{array}{ccc}
A_{\perp}\\
A_{\parallel}\\
a
\end{array}
\right)=0\,,
\end{equation}
with the mixing matrix
\begin{equation}  \label{mixmat}
{\cal M}=\left(
\begin{array}{ccc}
\Delta_{\perp} &\Delta_{\rm R}     & 0\\
\Delta_R       &\Delta_{\parallel} & \Delta_{a\gamma}\\
0              &\Delta_{a\gamma}   & \Delta_a\\
  \end{array}
  \right)\,,
\end{equation}
where
\begin{eqnarray}
\Delta_{a \gamma}&=&\gag
|{\bf B}_{\rm T}|/2 \,\ ,\\
\Delta_{\parallel, \perp} &=& \Delta_{\rm{pl}} +
\Delta_{\parallel, \perp}^{\rm CM} \,\ .
\end{eqnarray}
%%%%%%%%%%%%%%%%%%%%%%%%%%%%%%%%%%%%%%%%%%%%%%%%%%%%%%%%%%%%%%
In a plasma, the photons acquire an effective mass given by the
plasma frequency $\omega_{\rm pl}^2 = 4\pi\alpha\,n_e/m_e$ with $n_e$
the electron density, leading to
%%%%%%%%%%%%%%%%%%%%%%%%%%%%%%%%%%%%%%%%%%%%%%%%%%%%%%%%%%%%%%%%%%%%%
\begin{equation}
\Delta_{\rm pl} = - \frac{\omega_{\rm pl}^2}{2 \omega} \,\ .
\end{equation}
%%%%%%%%%%%%%%%%%%%%%%%%%%%%%%%%%%%%%%%%%%%%%%%%%%%%%%%%%%%%%%%%%%
Furthermore, the $\Delta^{\rm CM}_{\parallel, \perp}$ terms describe
the Cotton-Mouton effect, i.e.\ the birefringence of fluids in the
presence of a transverse magnetic field where
$|\Delta_{\parallel}^{\rm CM}-\Delta_{\perp}^{\rm CM}|\propto B_{\rm
T}^2$.  These terms are of little importance for the following
arguments and will thus be neglected. Finally, the Faraday rotation
term $\Delta_{\rm R}$, which depends on the energy and the
longitudinal component $B_z$, couples the modes $A_{\parallel}$ and
$A_{\perp}$.  While Faraday rotation is important when analyzing
polarized sources of photons, it plays no role for the problem at
hand.

With this simplification the $A_\perp$ component decouples, and the
propagation equations reduce to a 2-dimensional mixing problem with a
purely transverse field ${\bf B}={\bf B}_{\rm T}$
\begin{equation}
\left(\omega-i\partial_z +{\cal M}_2\right)
\left(
\begin{array}{cc}
A_\parallel\\a
\end{array}
\right)=0,
\end{equation}
with a 2-dimensional mixing matrix
\begin{equation}  \label{mixmat2}
{\cal M}_{2}=\left(
  \begin{array}{cc}
    \Delta_{\rm pl}&\Delta_{a \gamma}\\
    \Delta_{a \gamma}&\Delta_a
  \end{array}
  \right).
\end{equation}
The solution follows from diagonalization by the rotation angle
\begin{equation}
\label{tan}
\vartheta = \frac{1}{2}\arctan
\left(\frac{2\Delta_{a \gamma}}{\Delta_{\rm pl}-\Delta_a}\right).
\end{equation}
In analogy to the neutrino case~\cite{Kuo:1989qe}, the probability for
a photon emitted in the state $A_{\parallel}$ to convert into an axion
after travelling a distance $s$ is
\begin{eqnarray}\label{p1ga}
  P_0(\gamma\rightarrow a)&=&
  \left|\langle A_\parallel(0)|a(s)\rangle\right|^2\nonumber\\
  &=&\sin^2\left(2 \vartheta \right)
  \sin^2(\Delta_{\rm osc}s/2)\nonumber\\
  &=&\left(\Delta_{a \gamma} s\right)^2
  \frac{\sin^2(\Delta_{\rm osc} s /2)}
  {(\Delta_{\rm osc} s /2)^2} \;\ ,
\end{eqnarray}
where the oscillation wavenumber is given by
\begin{equation}\label{deltaosc}
\Delta_{\rm osc}^2=(\Delta_{\rm pl}-\Delta_a)^2 +
4 \Delta_{a \gamma}^2\,.
\end{equation}
The conversion probability is energy-independent when
$2|\Delta_{a\gamma}|\gg|\Delta_{\rm pl}-\Delta_{a}|$ or whenever the
oscillatory term in Eq.~(\ref{p1ga}) is small, i.e.\ $\Delta_{\rm osc}
s /2\ll1$, implying the limiting behavior $P_0=\left(\Delta_{a \gamma}
s\right)^2 \label{p1enind}$.

The propagation over many $B$-field domains is a truly 3-dimensional
problem, because different photon polarization states play the role of
$A_\parallel$ and $A_\perp$ in different domains.  This average is
enough to guarantee that the conversion probability over many domains
is an incoherent average over magnetic field configurations and photon
polarization states.  The probability after travelling over a distance
$r\gg s$, where $s$ is the domain size, is derived
in Appendix~A along the lines of Ref.~\cite{Grossman:2002by}
and is found to be
\begin{equation}\label{totpro}
P_{\gamma \to a}(r) = \frac{1}{3}
\left[1-\exp\left(-\frac{3P_0\,r}{2s}\right)\right]\,,
\end{equation}
with $P_0$ given by Eq.~(\ref{p1ga}).  As expected one finds that
for $r/s\to\infty$ the conversion probability saturates, so that on
average one third of all photons converts to axions.

%%%%%%%%%%%%%%%%%%%%%%%%%%%%%%%%%%%%%%%%%%%%%%%%%%%%%%%%%%%%%%%%%%%%%%
\section{Photon-axion conversion and supernova dimming}
\label{dimming}
%%%%%%%%%%%%%%%%%%%%%%%%%%%%%%%%%%%%%%%%%%%%%%%%%%%%%%%%%%%%%%%%%%%%%%

%%%%%%%%%%%%%%%%%%%%%%%%%%%%%%%%%%%%%%%%%%%%%%%%%%%%%%%%%%%%%%%%%%%%
\subsection{Observations}
%%%%%%%%%%%%%%%%%%%%%%%%%%%%%%%%%%%%%%%%%%%%%%%%%%%%%%%%%%%%%%%%%%%%%

In 1998, two groups using SNe~Ia as cosmic standard candles reported
first evidence for a luminosity-redshift relation that indicated that
the expansion of the universe was accelerating today~\cite{supnovR,
supnovP}.  The quantity relevant for SN~Ia observations is the
luminosity distance $d_L$ at redshift $z$, defined by
\begin{equation}
\label{distance}
d^2_L(z) = \frac{\mathcal{L}}{4 \pi \mathcal{F}}\,,
\end{equation}
where $\mathcal{L}$ is the absolute luminosity of the source and
$\mathcal{F}$ is the energy flux arriving at
Earth~\cite{supnovR,supnovP}.  In Friedmann-Robertson-Walker
cosmologies, the luminosity distance at a given redshift $z$ is a
function of the Hubble parameter $H_0$, the matter density
$\Omega_{\rm M}$, and the dark energy density $\Omega_\Lambda$.
Usually the data are expressed in terms of magnitudes
\begin{equation}
\label{magn}
m = M +  5 \log_{10} \left(\frac{d_L}{\rm  Mpc}\right)  + 25 \,,
\end{equation}
where $M$ is the absolute magnitude, equal to the value that $m$
would have at $d_L=10$ pc.

%%%%%%%%%%%%%%%%%%%%%%%%%%% FIGURE 2 %%%%%%%%%%%%%%%%%%%%%%%%%%%%%%%%%
\begin{figure}[t]
\centering
\includegraphics[height=8 cm]{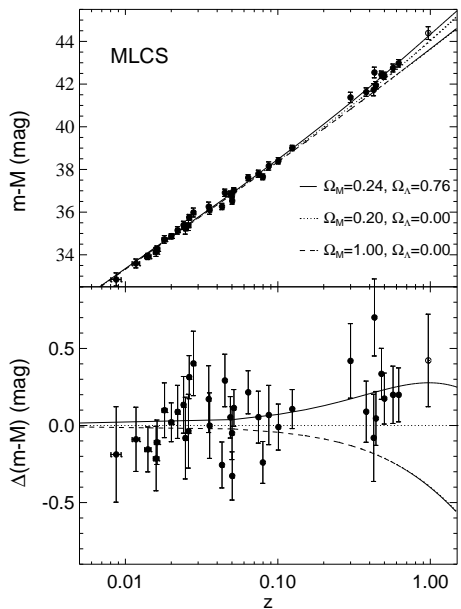}
\caption{\label{sn} \footnotesize\baselineskip=4mm SN~Ia Hubble diagram.  {\em Upper panel:}\/
Hubble diagram for low- and high-redshift SN~Ia samples. Overplotted
are three cosmologies: ``low'' and ``high'' $\Omega_{\rm M}$ with
$\Omega_\Lambda=0$ and the best fit for a flat cosmology
$\Omega_{\rm M}=0.24$ and $\Omega_\Lambda=0.76$.  {\em Lower
panel:}\/ Difference between data and models with $\Omega_{\rm
M}=0.20$ and $\Omega_\Lambda=0$. (Figure from Ref.~\cite{supnovR}
with permission.)}
\end{figure}
%%%%%%%%%%%%%%%%%%%%%%%%%%%%%%%%%%%%%%%%%%%%%%%%%%%%%%%%%%%%%%%%%%%%%%

Figure~\ref{sn} shows the Hubble diagram for SN~Ia samples at low
and high $z$.  The distances of high-redshift SNe are, on average,
$10\%$ to $15\%$ larger than in a low matter density ($\Omega_{\rm
M} =0.2$) Universe without dark energy ($\Omega_\Lambda =0$).
Therefore, objects of a fixed intrinsic brightness appear fainter if
the cosmic energy density budget is dominated by dark energy.  The
best fit of these data supports a Universe composed of a fraction of
dark matter $\Omega_{\rm M}\simeq 0.3$ and a fraction of dark energy
$\Omega_\Lambda \simeq 0.7$.

Dark energy has been associated with vacuum energy or an Einstein
cosmological constant that produces a constant energy density at all
times.  Defining the equation of state
%............................................................
\begin{equation}
w = \frac{p}{\rho} \,\ ,
\end{equation}
%...............................................................
the cosmological constant is characterized by $p=-\rho$, i.e.~$w=-1$.
From the Friedmann equations any component of the density budget with
equation of state $w < - 1/3$ causes cosmic acceleration.  SN~Ia data
imply that $w\gtrsim -0.5$ are disfavoured, supporting the cosmic
acceleration of the Universe~\cite{supnovP}.

\subsection{Interpretation in terms of photon-axion conversion}

To explore the effect of photon-axion conversion on SN dimming we
recast the relevant physical quantities in terms of natural
parameters.  The energy of optical photons is a few~eV. The strength
of widespread, all-pervading $B$-fields in the intergalactic medium
must be less than a few~$10^{-9}$~G over coherence lengths $s$ crudely
at the Mpc scale, according to the constraint from the Faraday
effect of distant radio sources~\cite{Kronberg:1993vk}.  Along a given
line of sight, the number of such domains in our Hubble radius is
about $N \approx H_0^{-1}/s\approx 4 \times 10^3$ for $s\sim 1$~Mpc.
The mean diffuse intergalactic plasma density is bounded by $n_e
\lesssim 2.7 \times 10^{-7}$~cm$^{-3}$, corresponding to the recent
WMAP measurement of the baryon density~\cite{Spergel:2003cb}.  Recent
results from the CAST experiment~\cite{Andriamonje:2004hi} give a
direct experimental bound on the axion-photon coupling of $\gag
\lesssim 1.16 \times 10^{-10}$~GeV$^{-1}$, comparable to the
long-standing globular-cluster limit~\cite{Raffeltbook}.  Suitable
representations of the mixing parameters are
\begin{eqnarray}  \label{eq12nm}
\frac{\Delta_{a \gamma}}{{\rm Mpc}^{-1}} &=&
0.15\;g_{10}\;B_{\rm nG}
\,,\nonumber\\
\frac{\Delta_a} {{\rm Mpc}^{-1}} &=&
-7.7 \times 10^{28} \left(\frac{m_a}{1\,\rm eV}\right)^2
\left(\frac{\omega}{1\,\rm eV}\right)^{-1}
\,,\nonumber\\
\frac{\Delta_{\rm pl}}{{\rm Mpc}^{-1}}  &=&
-11.1 \left(\frac{\omega}{1\,\rm eV}\right)^{-1}
\left(\frac{n_e}{10^{-7}\,{\rm cm}^{-3}}\right)\,,
\end{eqnarray}
where we have introduced $g_{10}=\gag/10^{-10}$ GeV$^{-1}$ and
$B_{\rm nG}$ is the magnetic field strength in nano-Gauss.

The mixing angle defined in Eq.~(\ref{tan}) is too small to yield a
significant conversion effect for the allowed range of axion masses
because $|\Delta_a| \gg |\Delta_{a \gamma}|$, $|\Delta_{\rm pl}|$.
Therefore, to ensure a sufficiently large mixing angle one has to
require nearly massless pseudoscalars, sometimes referred to as
``arions''~\cite{{Anselm:1981aw},{Anselm:1982ip}}.  For such
ultra-light axions a stringent limit from the absence of
$\gamma$-rays from SN~1987A gives $\gag\lesssim 1\times
10^{-11}$~GeV$^{-1}$~\cite{Brockway:1996yr} or even $\gag \lesssim
3\times 10^{-12}$~GeV$^{-1}$~\cite{Grifols:1996id}.  Henceforth we
will consider the pseudoscalars to be effectively massless, so that
our remaining independent parameters are $g_{10}B_{\rm nG}$ and
$n_e$. Note that $m_a$ only enters the equations via the term $m_a^2
- \omega_{\rm pl}^2$, so that for tiny but non-vanishing values of
$m_a$, the electron density should be interpreted as $n_{e,{\rm
eff}}=|n_e - m_a^2 m_e/(4 \pi \alpha)|$.

\begin{figure}[t]
\centering
\includegraphics[height=5 cm]{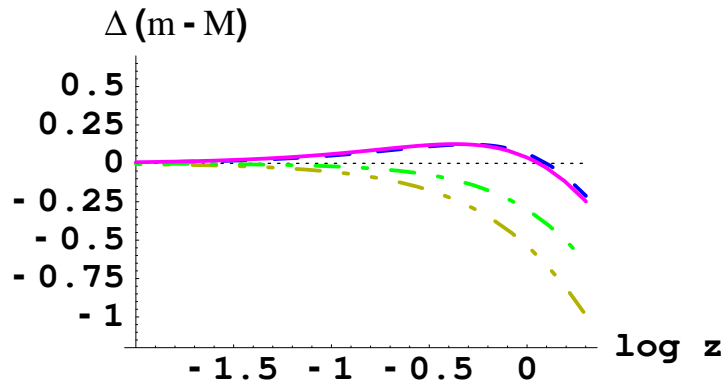}
\caption{\label{csaki} \footnotesize\baselineskip=4mm Hubble diagram for SNe~Ia for different
 cosmological models, relative to the curve with $\Omega_{\rm{tot}}
 =0$ (dotted horizontal line).  The dashed curve is a best fit to the
 SN data assuming that the Universe is accelerating ($\Omega_{\rm M}=0.3$,
 $\Omega_\Lambda=0.7$); the solid line is the photon-axion oscillation
 model with $\Omega_{\rm M}=0.3$ and $\Omega_S=0.7$, the dot-dashed line is
 $\Omega_{\rm M}=0.3$, $\Omega_S=0.7$ with no oscillation, the
 dot-dot-dashed line is for $\Omega_{\rm M}=1$ and again no
 oscillation. (Figure from Ref.~\cite{Csaki:2001yk} with permission.)}
\end{figure}

Allowing for the possibility of photon-axion oscillations in
intergalactic magnetic fields, the number of
photons emitted by the source and thus the flux $\mathcal{F}$ is
reduced to the fraction $P_{\gamma\to\gamma}=1-P_{\gamma\to a}$.
Therefore, the luminosity distance [Eq.~(\ref{distance})] becomes
\begin{equation}
d_L \to d_L/P_{\gamma \to \gamma}^{1/2} \,\ ,
\end{equation}
and the brightness [Eq.~(\ref{magn})]
\begin{equation}\label{isodim}
m \to m - \frac{5}{2}\log_{10}(P_{\gamma \to \gamma})\,.
\end{equation}
Distant SNe~Ia would eventually saturate ($P_{\gamma \to
\gamma}=2/3$), and hence they would appear $(3/2)^{1/2}$ times
farther away than they really are.  This corresponds to a maximum
dimming of approximately 0.4~mag.  Cs\'aki, Kaloper and Terning
(CKT~I) showed that if photon-axion conversion takes place, this
mechanism can reproduce the SN Hubble diagram~\cite{Csaki:2001yk},
assuming, for example, a nonstandard dark energy component
$\Omega_{S}=0.7$ with equation of state $w=-1/3$, which does not
produce cosmic acceleration (Fig.~\ref{csaki}).

%%%%%%%%%%%%%%%%%%%%%%%%%%% FIGURE 3 %%%%%%%%%%%%%%%%%%%%%%%%%%%%%%%%%
%%%%%%%%%%%%%%%%%%%%%%%%%%%%%%%%%%%%%%%%%%%%%%%%%%%%%%%%%%%%%%%%%%%%%%

%%%%%%%%%%%%%%%%%%%%%%%%%%% FIGURE 4 %%%%%%%%%%%%%%%%%%%%%%%%%%%%%%%%%
\begin{figure}[t]
\centering
\includegraphics[height=5 cm]{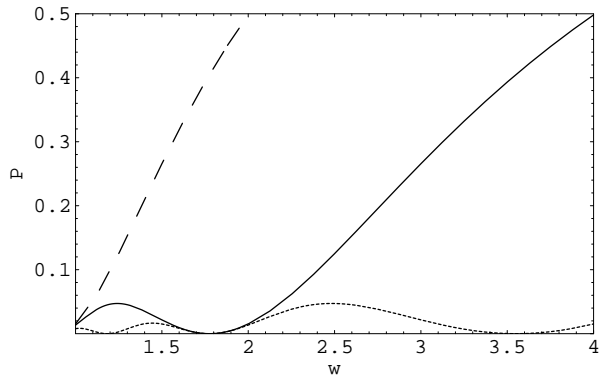}
 \caption{\label{deffayet} \footnotesize\baselineskip=4mm Ratio of the probability of conversion of
 photons to axions including the effects of the intergalactic plasma
 ($n_e\approx 10^{-7}{\rm cm}^{-3}$) and the probability of
 oscillations when this effect is not considered, as a function of the
 photon energy~$\omega$.  The curves are drawn for different size $s$
 of the magnetic domains: 0.5~Mpc (dashed line), 1~Mpc (solid line)
 and 2~Mpc (dotted line). (Figure from Ref.~\cite{Deffayet:2001pc} with permission.)}
\end{figure}
%%%%%%%%%%%%%%%%%%%%%%%%%%%%%%%%%%%%%%%%%%%%%%%%%%%%%%%%%%%%%%%%%%%%%%

However, in the model of CKT~I, plasma density effects were
neglected ($n_e=0$).  Later it was recognized that the conclusions
of CKT~I can be significantly modified when the effects of the
intergalactic plasma on the photon-axion oscillations are taken into
account~\cite{Deffayet:2001pc}.  In the presence of plasma effects,
the probability of oscillation is lower than before and it is no
longer achromatic (Fig.~\ref{deffayet}). SN observations not only
require dimming, but also that the dimming is achromatic. In fact,
SN observations put a constraint on the color excess between the $B$
and $V$ bands,
\begin{equation}
E[B-V]\equiv-2.5\ \ \log_{10}\left[{F^o(B) \over F^e(B)}\ \
{F^e(V) \over F^o(V)}\right],
\end{equation}
where $F^o$ or $F^e$ is the observed or emitted flux, respectively.
The $B$ and $V$ band correspond to $0.44~\mu {\rm m}$ and $0.55~\mu
{\rm m}$, respectively.  Observations constrain $E[B-V]$ to be lower
than $0.03$~\cite{supnovP}.  This can be translated to
\begin{equation}
\label{chrom}
P(\gamma \rightarrow a)_V \left[ {P(\gamma \rightarrow a)_B\over
P(\gamma
\rightarrow a)_V} -1 \right] < 0.03 \,\ .
\end{equation}
Therefore, assuming an electron density $n_e\approx n_{\rm baryons}=
n_{\gamma}\eta \sim 10^{-7}{\rm cm}^{-3}$, the model is ruled out in
most of the parameter space because of either an excessive photon
conversion or a chromaticity of the dimming~\cite{Deffayet:2001pc}.
Only fine-tuned parameters for the statistical properties of the
extragalactic magnetic fields would still allow this explanation.

On the other hand, Cs\'aki, Kaloper and Terning~\cite{Csaki:2001jk}
(CKT~II) criticized the assumed value of $n_e$ as being far too
large for most of the intergalactic space, invoking observational
hints for a value at least one order of magnitude smaller. As a
consequence, for values of $\omega_{\rm pl} \lesssim 6 \times
10^{-15}$~eV, corresponding to $n_e \lesssim 2.5 \times
10^{-8}$~cm$^{-3}$, one finds $|P_V - P_B| < 0.03$ so that the
chromaticity effect disappears very rapidly, and becomes
undetectable by present observations.

\begin{figure}[h]
\centering
\includegraphics[height=7cm]{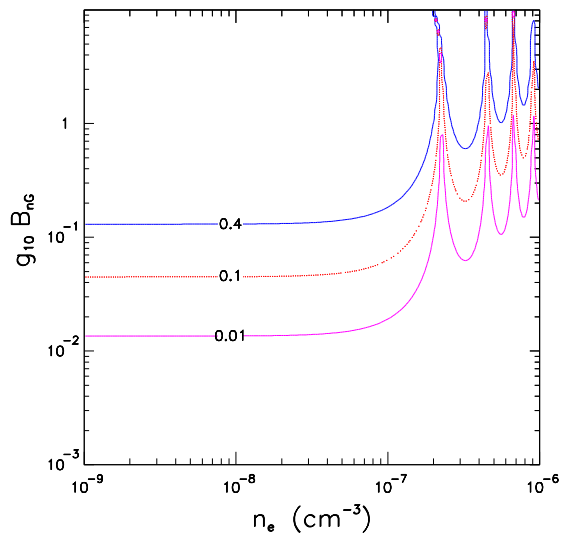}
 \caption{\label{fig1} \footnotesize\baselineskip=4mm
   Iso-dimming curves for an attenuation of 0.01, 0.1, and
   0.4~magnitudes. The photon energy of $4.0$~eV is representative of
   the B-band. The size of a magnetic domain is $s=1$~Mpc.
   (Figure from Ref.~\cite{Mirizzi:2005ng}.)}
\end{figure}

Figure~\ref{fig1} shows qualitatively the regions of $n_e$ and
$g_{10}B_{\rm nG}$ relevant for SN dimming at cosmological
distances.  To this end we show iso-dimming contours obtained from
Eq.~(\ref{isodim}) for a photon energy 4.0~eV and a magnetic domain
size $s=1$~Mpc.  For simplicity we neglect the redshift evolution of
the intergalactic magnetic field $B$, domain size $s$, plasma
density $n_e$, and photon frequency~$\omega$. Our iso-dimming curves
are intended to illustrate the regions where the photon-axion
conversion could be relevant. In reality, the dimming should be a
more complicated function since the intergalactic medium is expected
to be very irregular: there could be voids of low $n_e$ density, but
there will also be high density clumps, sheets and filaments and
these will typically have higher $B$ fields as well. However, the
simplifications used here are consistent with the ones adopted in
CKT~II and do not alter our main results.

The iso-dimming contours are horizontal in the low-$n_e$ and
low-$g_{10}B_{\rm nG}$ region.  They are horizontal for any
$g_{10}B_{\rm nG}$ when $n_e$ is sufficiently low.  From the
discussion in Sec.~\ref{sec:conversion} we know that the
single-domain probability $P_0$ of Eq.~(\ref{p1ga}) is indeed energy
independent when $|\Delta_{\rm osc} s| \ll 1$, i.e.\ for
$|\Delta_{\rm pl}| s/2\ll 1$ and $|\Delta_{a \gamma}| s \ll 1$. When
$n_e\lesssim {\rm few}~10^{-8}$~cm$^{-3}$ and $g_{10}B_{\rm nG}
\lesssim 4$, we do not expect an oscillatory behavior of the
probability. This feature is nicely reproduced by our iso-dimming
contours.  From Fig.~\ref{fig1} we also deduce that a significant
amount of dimming is possible only for $g_{10}B_{\rm nG}\gtrsim
4\times 10^{-2}$.

In CKT~I, where the effect of $n_e$ was neglected, a value $m_a \sim
10^{-16}$ eV was used.  In terms of our variables, this corresponds
to $n_{e,\rm eff} \approx 6 \times 10^{-12}$~cm$^{-3}$. As noted in
CKT~II, when plasma effects are taken into account, any value
$n_e\lesssim 2.5 \times 10^{-8}$~cm$^{-3}$ guarantees the required
achromaticity of the dimming below the 3\% level between the $B$ and
$V$~bands.  The choice $B_{\rm nG}$ of a few and $g_{10}\approx 0.1$
in CKT~I and~II falls in the region where the observed SN dimming
could be explained while being marginally compatible with the bounds
on the intergalactic $B$ field and on the axion-photon
coupling~$g_{10}$.

%%%%%%%%%%%%%%%%%%%%%%%%%%% FIGURE 5 %%%%%%%%%%%%%%%%%%%%%%%%%%%%%%%%%
%%%%%%%%%%%%%%%%%%%%%%%%%%%%%%%%%%%%%%%%%%%%%%%%%%%%%%%%%%%%%%%%%%%%%%

%%%%%%%%%%%%%%%%%%%%%%%%%%%%%%%%%%%%%%%%%%%%%%%%%%%%%%%%%%%%%%%%%%%%%%
\section{CMB Constraints}                                 \label{CMBC}
%%%%%%%%%%%%%%%%%%%%%%%%%%%%%%%%%%%%%%%%%%%%%%%%%%%%%%%%%%%%%%%%%%%%%%

If photon-axion conversion over cosmological distances is
responsible for the SN~Ia dimming, the same phenomenon should also
leave an imprint in the CMB.  A similar argument was previously
considered for photon-graviton conversion~\cite{Chen:1994ch}.
Qualitatively, in the energy-dependent region of $P_{\gamma\to a}$
one expects a rather small effect due to the low energy of CMB
photons ($\omega \sim 10^{-4}$~eV). However, when accounting for the
incoherent integration over many domains crossed by the photon,
appreciable spectral distortions may arise in view of the accuracy
of the CMB data at the level of one part in $10^{4}$--$10^{5}$. For
the same reason, in the energy-independent region, at much lower
values of $n_e$ than for the SNe~Ia, the constraints on
$g_{10}B_{\rm nG}$ are expected to be quite severe.  The depletion
of CMB photons in the patchy magnetic sky and its effect on the CMB
anisotropy pattern have been previously
considered~\cite{Csaki:2001yk}.  However, more stringent limits come
from the distortion of the overall blackbody
spectrum~\cite{Mirizzi:2005ng}.

%%%%%%%%%%%%%%%%%%%%%%%%%%% FIGURE 6 %%%%%%%%%%%%%%%%%%%%%%%%%%%%%%%%%
\begin{figure}[b]
\centering
\includegraphics[height=7cm, angle=90]{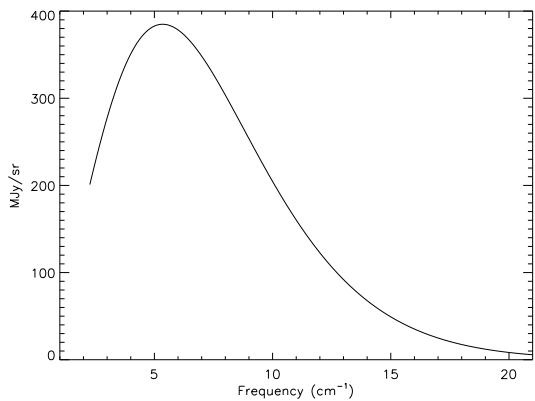}
 \caption{\label{cmb} \footnotesize\baselineskip=4mm Uniform CMB spectrum and fit to the
  blackbody spectrum. Uncertainties are a small fraction of the
  line thickness.
  (Figure from Ref.~\cite{Fixsen:1996nj} with permission.)}
\end{figure}
%%%%%%%%%%%%%%%%%%%%%%%%%%%%%%%%%%%%%%%%%%%%%%%%%%%%%%%%%%%%%%%%%%%%%%

To this end the COBE/FIRAS data for the experimentally measured
spectrum were used, corrected for foregrounds~\cite{Fixsen:1996nj}.
Note that the new calibration of FIRAS~\cite{Mather:1998gm} is
within the old errors and would not change any of our conclusions.
The $N = 43$ data points $\Phi^{\rm exp}_i$ at different energies
$\omega_i$ are obtained by summing the best-fit blackbody spectrum
(Fig.~\ref{cmb}) to the residuals reported in
Ref.~\cite{Fixsen:1996nj}. The experimental errors $\sigma_i$ and
the correlation indices $\rho_{ij}$ between different energies are
also available.  In the presence of photon-axion conversion, the
original intensity of the ``theoretical blackbody'' at temperature
$T$
\begin{equation}
\label{planck}
\Phi^0({\omega},T) = \frac{\omega^3}{ 2 \pi^2}
\big[ \exp (\omega/T )-1 \big]^{-1}
\end{equation}
would convert to a deformed spectrum that is given by
\begin{equation}
\Phi({\omega},T)=\Phi^0({\omega},T)P_{\gamma\to\gamma}({\omega}) \,\ .
\end{equation}
In Ref.~\cite{Mirizzi:2005ng}, we  build the reduced chi-squared
function
\begin{equation}
\chi_\nu^2(T,\lambda)=\frac{1}{{N}-1}
\sum_{i,j=1}^{N} {\Delta \Phi_i} (\sigma^2)^{-1}_{ij}
{\Delta \Phi_j} \,,
\end{equation}
where
\begin{equation}
\Delta \Phi_i = \Phi^{\rm exp}_i-\Phi^0({\omega}_i,T)
P_{\gamma\to\gamma}({\omega_i},\lambda)
\end{equation}
is the $i$-th residual, and
\begin{equation}
\sigma^2_{ij}= \rho_{ij} \sigma_{i} \sigma_{j}
\end{equation}
is the covariance matrix. We minimize this function with respect to
$T$ for each point in the parameter space $\lambda=(n_e,g_{10}B_{\rm
nG})$, i.e.\ $T$ is an empirical parameter determined by the
$\chi_\nu^2$ minimization for each $\lambda$ rather than being fixed
at the standard value $T_0=2.725\pm0.002$~K \cite{Mather:1998gm}. In
principle, one should  marginalize also over the galactic foreground
spectrum~\cite{Fixsen:1996nj}. However, this is a subleading effect
relative to the spectral deformation caused by the photon-axion
conversion.

%%%%%%%%%%%%%%%%%%%%%%%%%%% FIGURE 7 %%%%%%%%%%%%%%%%%%%%%%%%%%%%%%%%%
\begin{figure}[t]
\centering
\includegraphics[height=7cm]{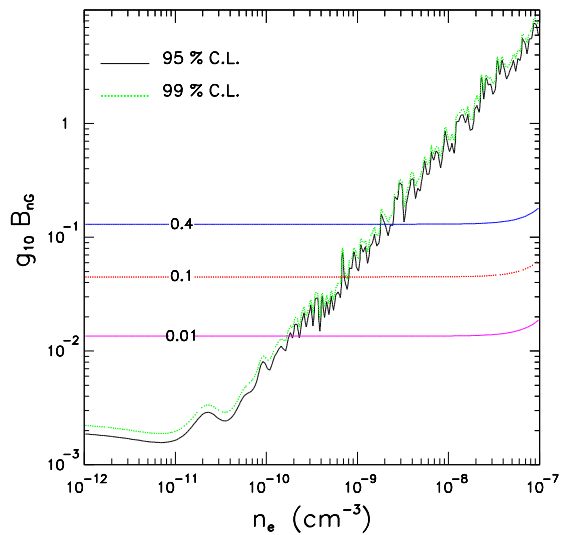}
\caption{\label{fig2} \footnotesize\baselineskip=4mm
 Exclusion plot for axion-photon conversion based
  on the COBE/FIRAS CMB spectral data.  The region above the solid
  curve is excluded at 95\% CL whereas the one above the dotted
  curve is excluded at 99\% CL. The size of each magnetic domain is
  fixed at $s=1$~Mpc. We also reproduce the iso-dimming contours from
  Fig.~\ref{fig1}. (Figure from Ref.~\cite{Mirizzi:2005ng}.)}
\end{figure}
%%%%%%%%%%%%%%%%%%%%%%%%%%%%%%%%%%%%%%%%%%%%%%%%%%%%%%%%%%%%%%%%%%%%%%

In Fig.~\ref{fig2} we show the exclusion contour in the plane of
$n_e$ and $g_{10}B_{\rm nG}$. The region above the continuous curve
is the excluded region at 95\% CL, i.e.\ in this region the chance
probability to get larger values of $\chi_\nu^2$ is lower than~5\%.
We also show the corresponding 99\% CL contour which is very close
to the 95\% contour so that another regression method and/or
exclusion criterion would not change the results very much.  Within
a factor of a few, the same contours also hold if one varies the
domain size $s$ within a factor of~10. Comparing this exclusion plot
with the iso-dimming curves of Fig.~\ref{fig1} we conclude that the
entire region $n_e \lesssim 10^{-9}$~cm$^{-3}$ is excluded as a
leading explanation for SN~dimming.

A few comments are in order.  Intergalactic magnetic fields probably
are a relatively recent phenomenon in the cosmic history, arising
only at redshifts of a few. As a first approximation we have then
considered the photon-axion conversion as happening for present
($z=0$) CMB photons.  Since $P_{\gamma\to \gamma}$ is an increasing
function of the photon energy $\omega$, our approach leads to
conservative limits.  Moreover, we assumed no correlation between
$n_e$ and the intergalactic magnetic field strength. It is however
physically expected that the fields are positively correlated with
the plasma density so that relatively high values of $g_{10}B_{\rm
nG}$ should be more likely when $n_e$ is larger.  Our constraints in
the region of $n_e\gtrsim 10^{-10}$~cm$^{-3}$ are thus probably
tighter than what naively appears.

%%%%%%%%%%%%%%%%%%%%%%%%%%%%%%%%%%%%%%%%%%%%%%%%%%%%%%%%%%%%%%%%%%%%%%
\section{QSO Constraints}                                  \label{QSO}
%%%%%%%%%%%%%%%%%%%%%%%%%%%%%%%%%%%%%%%%%%%%%%%%%%%%%%%%%%%%%%%%%%%%%%

%%%%%%%%%%%%%%%%%%%%%%%%%%% FIGURE 8 %%%%%%%%%%%%%%%%%%%%%%%%%%%%%%%%%
\begin{figure}[t]
\centering
\includegraphics[height=7cm]{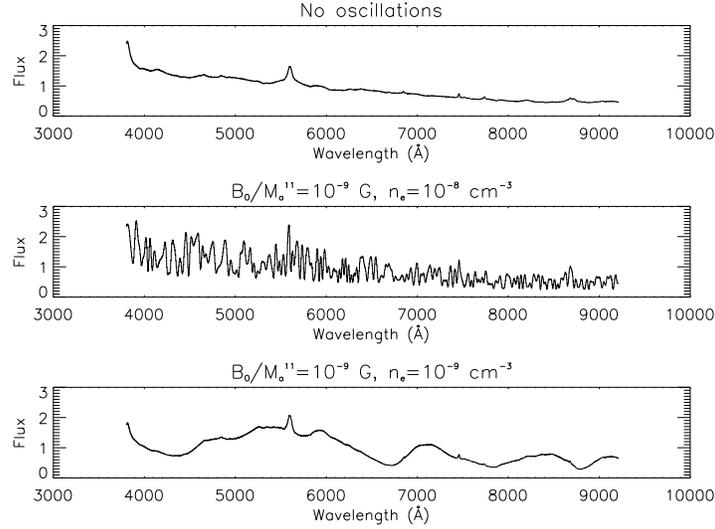}
\caption{\label{qso} \footnotesize\baselineskip=4mm
  Simulated quasar spectra at $z=1$ for different
photon-axion oscillation scenarios. (Figure from
Ref.~\cite{Ostman:2004eh} with permission.)}
\end{figure}
%%%%%%%%%%%%%%%%%%%%%%%%%%%%%%%%%%%%%%%%%%%%%%%%%%%%%%%%%%%%%%%%%%%%%%
%\vskip1cm

CMB limits are nicely complementary to the ones obtained from the
effects of photon-axion conversion on quasar colors and
spectra~\cite{Ostman:2004eh}.  One effect of photon-axion
oscillations is that a dispersion is added to the quasar spectra due
to the energy dependence of the effect. By comparing the dispersion
in observed in quasar spectra with the dispersion in simulated ones,
one can find out if the model behind each simulation is allowed. The
SuperNova Observation Calculator (SNOC)~\cite{Goobar:2002vm} was
used~\cite{Ostman:2004eh} to simulate the effects of photon-axion
oscillations on quasar observations (Fig.~\ref{qso}). If the
simulated dispersion is smaller than the observed, one cannot
exclude the scenario since real quasars have an intrinsic
dispersion.

\begin{figure}[t]
\centering
\includegraphics[height=7cm]{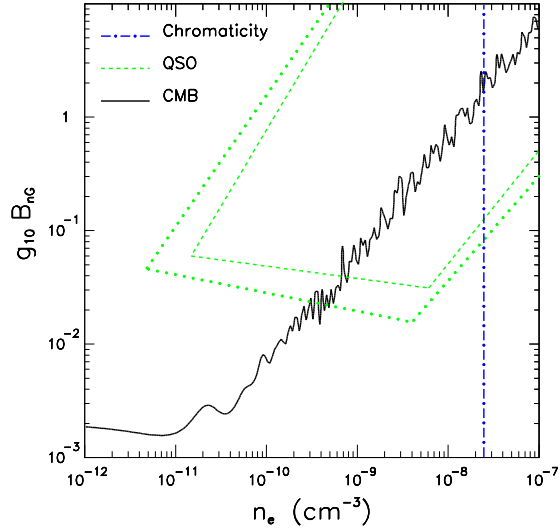}
 \caption{\label{fig3} \footnotesize\baselineskip=4mm
  Exclusion plot for photon-axion conversion.  The
   region to the right of the dot-dashed line is excluded by requiring
   achromaticity of SN~Ia dimming. The region inside the dashed lines
   is excluded by the dispersion in QSO spectra. Assuming an intrinsic
   dispersion of 5\% in QSO spectra, the excluded region could be
   extended up to the dotted curve. The CMB argument excludes the
   entire region above the solid curve at 95\% CL. (Figure
   from Ref.~\cite{Mirizzi:2005ng}.)}
\end{figure}

In Fig.~\ref{fig3} we superimpose the
CMB exclusion contours with the schematic region excluded by quasars
%%%%%%%%%%%%%%%%%%%%%%%%%%%%%%%%%%%%%%%%%%%%%%%%%%%%%%%%%%%%%%%%%%%%%%
\footnote{We use the exclusion regions of astro-ph/0410501v1.
  In the published version~\cite{Ostman:2004eh}, corresponding to
  astro-ph/0410501v2, the iso-dimming curves were erroneously changed.
  The difference is that in version~1 the angle $\alpha$ in Eq.~(3) of
  Ref.~\cite{Ostman:2004eh} that characterizes the random magnetic
  field direction was correctly taken in the interval
  0--360$^\circ$ whereas in version 2 it was taken in the interval
  0--90$^\circ$ (private communication by the authors).}.
%%%%%%%%%%%%%%%%%%%%%%%%%%%%%%%%%%%%%%%%%%%%%%%%%%%%%%%%%%%%%%%%%%%%%%
The region to the right of the dot-dashed line is excluded by
requiring achromaticity of SN~Ia dimming~\cite{Csaki:2001jk}. The
region inside the dashed lines is excluded by the dispersion in QSO
spectra. Moreover, assuming an intrinsic dispersion of 5\% in these
spectra, the excluded region could be enlarged up to the dotted
lines. The CMB argument excludes the region above the solid curve at
95\%~CL.

A cautionary remark is in order when combining the two constraints.
As we have discussed in the previous section, CMB limits on
photon-axion conversion are model independent. On the other hand,
the limits placed by the QSO spectra may be subject to loop holes,
since they are based on a full correlation between the intergalactic
electron density and the magnetic field strength, which is
reasonable but not well established observationally.

%%%%%%%%%%%%%%%%%%%%%%%%%% FIGURE 8 %%%%%%%%%%%%%%%%%%%%%%%%%%%%%%%%%
%%%%%%%%%%%%%%%%%%%%%%%%%%%%%%%%%%%%%%%%%%%%%%%%%%%%%%%%%%%%%%%%%%%%%

%%%%%%%%%%%%%%%%%%%%%%%%%%%%%%%%%%%%%%%%%%%%%%%%%%%%%%%%%%%%%%%%%%%%%%
\section{Constraints from angular diameter distance}
%%%%%%%%%%%%%%%%%%%%%%%%%%%%%%%%%%%%%%%%%%%%%%%%%%%%%%%%%%%%%%%%%%%%%%

We now turn briefly to two other types of constraint on the
photon-axion conversion mechanism.  The first is based on angular
diameter distance measurements of radio-galaxies. For a source of
linear radius $r$ and angular diameter $\theta$, the angular
diameter distance is
\begin{equation}
d_A = \frac{2r}{\theta} \,\ .
\end{equation}
In metric theories where photons travel on null geodesics and their
number is conserved, the angular distance $d_A$ and the luminosity
distance $d_L$ are fundamentally related by the reciprocity
relation~\cite{SEF}
\begin{equation}
d_L(z) = (1+z)^2 d_A(z) \,\ .
\end{equation}
Photon-axion conversion in intergalactic magnetic fields would not
affect the angular-diameter distance~\cite{bakuI, bakuII} and hence
would cause a fundamental asymmetry between measurements of $d_L(z)$
and $d_A(z)$.

In a first search for a violation of the reciprocity relation, a
joint analysis of high-redshift SNe~Ia [$d_L(z)$] and radio galaxies
[$d_A(z)$] was undertaken~\cite{bakuI}.  The results do not favour
the loss of photons and hence disfavour mixing.  However, this
constraint is less robust than the QSO one because it is affected by
possibly large systematic errors that are difficult to
quantify~\cite{uam}.

Since angular-diameter distance is immune to the loss of photons,
the axion-conversion versus accelerating-universe ambiguity in the
interpretation can be resolved~\cite{Song:2005af} by combining CMB
acoustic peak measurements with the recent detection of baryon
oscillations in galaxy power spectra~\cite{Eisenstein:2005su}.  This
combination excludes a non-accelerating dark-energy species at the
$4\sigma$ level regardless of the level of the axion coupling.

%%%%%%%%%%%%%%%%%%%%%%%%%%%%%%%%%%%%%%%%%%%%%%%%%%%%%%%%%%%%%%%%%%%%%%
\section{Conclusions}                              \label{conclusions}
%%%%%%%%%%%%%%%%%%%%%%%%%%%%%%%%%%%%%%%%%%%%%%%%%%%%%%%%%%%%%%%%%%%%%%

We have reviewed the intriguing and phenomenologically
rich~\cite{Das:2004qk} mechanism of conversion of photons into very
low-mass axion-like particles in the presence of intergalactic
magnetic fields. We have examined the existing astrophysical and
cosmological limits on this model, coming from the distortion of the
CMB spectrum, from the quasar dispersion, and from the angular
diameter distance, including the baryon oscillations detected in
large-scale structure surveys.

In particular, we have shown that the resulting CMB spectral
deformation excludes a previously allowed parameter region
corresponding to very low densities of the intergalactic medium
(IGM). These limits are complementary to the ones derived from QSO
dispersion which place serious constraints on the axion-photon
conversion mechanism, especially for relatively large densities of the
IGM.  As a result, it appears that the photon-axion conversion will
not play a leading role for the apparent SN~Ia dimming.

It may still happen that ultra-light or massless axions play an
important cosmological role.  For example, it was shown that by
adding a photon-axion conversion mechanism on top of a dark energy
model with $w\gtrsim -1$, one can mimic cosmic equations of state as
negative as $w\simeq- 1.5$~\cite{Csaki:2004ha}.  Although at present
there is no need for such an extreme equation of state, it is an
interesting possibility to keep in mind, especially since
alternative explanations as ghost/phantom fields usually pose a
threat to very fundamental concepts in general relativity and
quantum field theory.

%%%%%%%%%%%%%%%%%%%%%%%%%%%%%%%%%%%%%%%%%%%%%%%%%%%%%%%%%%%%%%%%%%%%%%
%% Acknowledgments %%%%%%%%%%%%%%%%%%%%%%%%%%%%%%%%%%%%%%%%%%%%%%%%%%%
%%%%%%%%%%%%%%%%%%%%%%%%%%%%%%%%%%%%%%%%%%%%%%%%%%%%%%%%%%%%%%%%%%%%%%
\section*{Acknowledgments}

A.~M.\ and G.~R.\ thank the organizers of the Joint ILIAS-CAST-CERN
Axion Training at CERN for their kind hospitality.  In Munich, this
work was supported, in part, by the Deutsche Forschungsgemeinschaft
under grant No.~SFB 375 and by the European Union under the ILIAS
project, contract No.~RII3-CT-2004-506222. A.M.\ is supported, in
part, by the Istituto Nazionale di Fisica Nucleare (INFN) and by the
Ministero dell'Istruzione, Universit\`a e Ricerca (MIUR) through the
``Astroparticle Physics'' research project.

%%%%%%%%%%%%%%%%%%%%%%%%%%%%%%%%%%%%%%%%%%%%%%%%%%%%%%%%%%%%%%%%%%%%%%
\appendix
%%%%%%%%%%%%%%%%%%%%%%%%%%%%%%%%%%%%%%%%%%%%%%%%%%%%%%%%%%%%%%%%%%%%%%

%%%%%%%%%%%%%%%%%%%%%%%%%%%%%%%%%%%%%%%%%%%%%%%%%%%%%%%%%%%%%%%%%%%%%%
\section{Photon-axion conversion in a random background}
\label{averageprob}
%%%%%%%%%%%%%%%%%%%%%%%%%%%%%%%%%%%%%%%%%%%%%%%%%%%%%%%%%%%%%%%%%%%%%%

In the following, we derive Eq.~(\ref{totpro}) along the lines of
Ref.~\cite{Grossman:2002by}.  It is assumed that photons and axions
traverse $N$ domains of equal length $s$.  The component of the
magnetic field perpendicular to the direction of flight ${\vec
B}_{\rm T}$ is constant within each domain and of equal strength
($B\equiv|{\vec B}_{\rm T}|$) in each domain, but it is assumed to
have a random orientation in each cell.

We begin with an initial state that is a coherent superposition of an
axion and the two photon states $|A_{1,2} \rangle$ that correspond
respectively to photons polarized parallel and perpendicular to the
magnetic field in the first domain,
%%%%%%%%%%%%%%%%%%%%%%%%%%%%%%%%%%%%%%%%%%%%%%%%%%%%%
\begin{equation}
\kappa_1(0) |A_1\rangle
+\kappa_2(0)|A_2\rangle +\kappa_a(0)|a\rangle \,\ .
\end{equation}
%%%%%%%%%%%%%%%%%%%%%%%%%%%%%%%%%%%%%%%%%%%%%%%%%%%%%%%%%%
The initial photon and axion fluxes are
%%%%%%%%%%%%%%%%%%%%%%%%%%%%%%%%%%%%%%%%%%%%%%%%%%%%%%%%%%%%
\begin{eqnarray}
I_\gamma(0)&\sim& |\kappa_1(0)|^2 +|\kappa_2(0)|^2 \,\ , \\
I_a(0)&\sim & |\kappa_a(0)|^2 \,\ ,
\end{eqnarray}
%%%%%%%%%%%%%%%%%%%%%%%%%%%%%%%%%%%%%%%%%%%%%%%%%%%%%%%%%%%%
respectively. In the $n$-th domain the transverse magnetic field
${\vec B}_{\rm T}$ is tilted by an angle $\gamma_n$ compared to the
magnetic field in the first domain
%%%%%%%%%%%%%%%%%%%%%%%%%%%%%%%%%%%%%%%%%%%%%%%%%%%%%%%%%%%%
\begin{eqnarray}
|A_\parallel^n\rangle &=& c_n|A_1\rangle+s_n|A_2\rangle \,\ , \\
|A_\perp^n\rangle &=& -s_n|A_1\rangle+c_n|A_2\rangle \,\ ,
\end{eqnarray}
%%%%%%%%%%%%%%%%%%%%%%%%%%%%%%%%%%%%%%%%%%%%%%%%%%%%%%%%%%%%%%
or
\begin{eqnarray}
c_1(z) &=& c_n \kappa_\parallel^n(z)-s_n \kappa_\perp^n(z)\,\ , \\
c_2(z) &=& s_n \kappa_\parallel^n(z)+c_n \kappa_\perp^n(z)\,\ ,
\end{eqnarray}
%%%%%%%%%%%%%%%%%%%%%%%%%%%%%%%%%%%%%%%%%%%%%%%%%%%%%%%%%%%%%%%
where $c_n=\cos\gamma_n$ and $s_n=\sin\gamma_n$.  Only photons
polarized parallel to the magnetic field mix with axions.  The
values of the transition elements are equal in each domain since the
magnitude of the magnetic field $B$ has been assumed to be the same
everywhere. The transition probability $P_0$ for photon to axion
oscillation in one domain is given by Eq.~(\ref{p1ga}), and the
photon survival probability is $1-P_0$.  At the end of the $n$-th
domain, the photon and axion fluxes are
\bea I_\gamma(n+1)&\sim& (1-P_0c_n^2)|\kappa_1(z_n)|^2\\
&&+(1-P_0s_n^2)|\kappa_2(z_n)|^2+P_0|\kappa_a(z_n)|^2+\dots
\nonumber\\
I_a(n+1)&\sim& P_0c_n^2|\kappa_1(z_n)|^2\\
&&+P_0s_n^2|\kappa_2(z_n)|^2+(1-P_0)|\kappa_a(z_n)|^2+\dots
\nonumber \eea where the dots represent terms that are proportional
to $c_n$, $s_n$, or $c_n s_n$. We have defined $z_n=(n-1)s$. The
coefficients $\kappa_{1}$, $\kappa_{2}$ and $\kappa_{a}$are taken at
the beginning of the $n$-th domain.

Next, we assume that the transition probability in one domain is
small, i.e.~$P_0\ll 1$, and that the direction of the magnetic field
is random, i.e.~$\gamma_n$ is a random variable so that
$\gamma_{n+1}-\gamma_n$ is of order unity. Due to the randomness of
the magnetic field, in this limit $c_n^2$ and $s_n^2$ can be
replaced by their average value $1/2$, while the interference terms
$c_n$, $s_n$ and $c_n s_n$ are averaged to zero. Using
%%%%%%%%%%%%%%%%%%%%%%%%%%%%%%%%%%%%%%%%%%
\begin{eqnarray}
I_\gamma(n) &\sim & |\kappa_1(z_n)|^2+|\kappa_2(z_n)|^2 \,\ , \\
I_a &\sim & |\kappa_a(z_n)|^2 \,\ ,
\end{eqnarray}
%%%%%%%%%%%%%%%%%%%%%%%%%%%%%%%%%%%%%%%%%%%%%%%%%%%%%%%%%%%%%%%%%%%%
one arrives at
\bea
\pmatrix{I_\gamma(n+1)\cr I_a(n+1)}&=&\pmatrix{1-\frac{1}{2}P_0
&P_0\cr \frac{1}{2}P_0&1-P_0}\pmatrix{I_\gamma(n)\cr I_a(n)}
\\
%\eea
%\bea
% \cr
&=&\frac{1}{3}\pmatrix{2+\left(1-\frac{3}{2}P_0\right)^{n+1}&
2-2\left(1-\frac{3}{2}P_0\right)^{n+1}\cr
1-\left(1-\frac{3 }{2}P_0\right)^{n+1}&
1+2\left(1-\frac{3 }{2}P_0\right)^{n+1}}
\pmatrix{I_\gamma(0)\cr I_a(0)}.\nonumber
\eea
As the number of domains is large one can
replace $\left(1-3P_0/2\right)^{n+1}$ with the limiting function
 $\exp\left[-3 P_0 z/(2s)\right]$ to arrive at the final expressions
\bea
I_\gamma(z)&=&I_\gamma(0)-P_{\gamma\to a}[I_\gamma(0)-2 I_a(0)] \,, \\
I_a(z)&=&I_a(0)+P_{\gamma\to a}[I_\gamma(0)-2 I_a(0)] \,,
\eea
with
\be
P_{\gamma\to a}=
\frac{1}{3}\left[1-\exp\left(-\frac{3 P_0 z}{2\,s}\right)\right]\,.
\ee

%\newpage

%%%%%%%%%%%%%%%%%%%%%%%%%%%%%%%%%%%%%%%%%%%%%%%%%%%%%%%%%%%%%%%%%%%%%%

%%%%%%%%%%%%%%%%%%%%%%%%%%%%%%%%%%%%%%%%%%%%%%%%%%%%%%%%%%%%%%%%%%%%%%

%%%%%%%%%%%%%%%%%%%%%%%%%%%%%%%%%%%%%%%%%%%%%%%%%%%%%%%%%%%%%%%%%%%%%%

%\clearpage

%%%%%%%%%%%%%%%%%%%%%%%%%%%%%%%%%%%%%%%%%%%%%%%%%%%%%%%%%%%%%%%%%%%%%%
%%%  Bibliography  %%%%%%%%%%%%%%%%%%%%%%%%%%%%%%%%%%%%%%%%%%%%%%%%%%%
%%%%%%%%%%%%%%%%%%%%%%%%%%%%%%%%%%%%%%%%%%%%%%%%%%%%%%%%%%%%%%%%%%%%%%

%%%%%%%%%%%%%%%%%%%%%%%%%%%%%%%%%%%%%%%%%%%%%%%%%%%%%%%%%%%%%%%%%%%%%%
\end{document}